hep-ph/0608052
\documentclass{article}
\usepackage{hiph-art}
\usepackage{graphicx}
\volnumber{?} \issuenumber{?} \edyear{?}                                 
\frompage{?} \topage{?}                                                  
\recrevdate{date}                                                        
\def\rightmark{Isentropic Equation of State of Two-Flavour QCD in a Quasi-Particle Model}
\def\leftmark{M. Bluhm, B. K\"ampfer, R. Schulze and D. Seipt}
 
\title{Isentropic Equation of State of Two-Flavour QCD\\ 
in a Quasi-Particle Model}
\authors{ 
{M. Bluhm$^1$, B. K\"ampfer$^{1,2}$, R. Schulze$^2$ and D. Seipt$^2$ %
\index{Bluhm, M.} 
\index{K\"ampfer, B.} 
\index{Schulze, R.} 
\index{Seipt, D.} 
}\\[2.812mm]
{\normalsize
\hspace*{-8pt}$^1$ Forschungszentrum Rossendorf, PF 510119, 01314 Dresden, Germany\\[0.2ex] 
\hspace*{-8pt}$^2$ Institut f\"ur Theoretische Physik, TU Dresden, 01062 Dresden, Germany
}}
 
\abstract{
  We examine the isentropic QCD equation of state within a quasi-particle model being adjusted to first 
  principle QCD calculations of two quark flavours. In particular, we compare with Taylor expansion 
  coefficients of energy and entropy densities and with the isentropic trajectories describing the hydrodynamical 
  expansion of a heavy-ion collision fireball. } 

 
\makeindex
\begin{document}
 
\maketitle

\section{Introduction}\label{intro}

With growing evidence, ideal hydrodynamics appears to be successfully 
describing the expansion stage of strongly interacting matter in relativistic heavy-ion collisions 
\cite{Kolb1,Kolb2,Shuryak-Teaney,Shu,Heinz}. The heart of hydrodynamics is the equation 
of state (EoS) which is needed as important interrelation among the state variables of strongly interacting 
matter in order to solve the hydrodynamic equations of motion. The EoS can, for instance, relate 
pressure, energy density and baryon density in the form $p=p(e,n_B)$. The influence of different model 
EoS on observables was analyzed, e.~g. in \cite{Shuryak-Teaney,HuovinenPRL,Huonew}. 
Recent progress in QCD calculations (performed numerically on a discretized space-time grid, dubbed lattice), 
though, allows the calculation of these EoS quantities 
from first principles. It is the aim of the present paper to compare our quasi-particle model with some 
available lattice QCD results. 

Our paper is organized as follows. In section \ref{sec:QPM} we review our quasi-particle model. 
In section \ref{sec:EoS}, we compare the model with recent two flavour lattice QCD results for pressure, 
energy density and entropy density. In particular, we focus on the isentropic trajectories and the 
EoS along those paths of constant entropy per baryon. The results are summarized in section \ref{sec:conclusions}. 
 
\section{Quasi-Particle Model \label{sec:QPM}} 

The QCD EoS was often formulated in terms of quasi-particles with effectively modified properties 
due to the strong interaction (cf. \cite{Levai,Rischke} and references therein). More recently developed 
quasi-particle models are proposed in \cite{Schneider,Letessier,Rebhan,Thaler,Ivanov,Khvo,Bannur}. In 
our quasi-particle model (QPM) \cite{Peshier}, we employ as thermodynamic potential the pressure $p$ in 
thermal equilibrium. Concentrating on the case of $N_f=2$ light quark flavours with one chemical 
potential $\mu_q$, $p$ as a function of temperature $T$ and $\mu_q$ reads 
\begin{equation}
  \label{e:pres0}
  p (T,\mu_q) = \sum_{a = q,g} p_a - B(T,\mu_q) 
\end{equation}
with partial pressures of quarks (q) and transverse gluons (g) \linebreak $p_a = d_a/(6 \pi^2) \int_0^\infty dk 
k^4\left( f_a^+ + f_a^- \right)/\omega_a$. 
Here, $d_q=2N_fN_c$, $d_g=N_c^2-1$, $N_c=3$, and $f_a^\pm = (\exp( [\omega_a \mp \mu_a]/ T) +S_a)^{-1}$ 
with $S_{q}=1$ for fermions and $S_g=-1$ for bosons. Note $\mu_g=0$. 

The quasi-particles propagate predominantly on-shell with a dispersion relation 
approximated by the asymptotic mass shell expression near the light cone, $\omega_a = \sqrt{k^2 + m_a^2}$, 
where $m_a^2 = m_{0;a}^2 + \Pi_a$ \cite{Pisarski} with self-energies $\Pi_a$ and $m_{0;g} = 0$. 
For $\Pi_a$, we employ the asymptotic expressions of the gauge 
independent hard-thermal loop / hard-density loop self-energies \cite{leBellac}. 
The mean field interaction term $B(T,\mu_q)$ in~(\ref{e:pres0}) is determined by thermodynamic 
self-consistency and stationarity of the thermodynamic potential under functional variation with respect 
to the self-energies, $\delta p / \delta \Pi_a = 0$~\cite{Gorenstein}. 

Replacing the running coupling $g^2$ entering $\Pi_a$ by an effective coupling $G^2$ depending on $T$ and $\mu_q$, 
non-perturbative effects within the strongly interacting system are accommodated in our model. 
A convenient parametrization of $G^2(T,\mu_q=0)$ (cf. \cite{Bluhm04p,Blu05}) is 
\begin{equation}
  \label{e:G2param}
  G^2(T,\mu_q=0) = \left\{
    \begin{array}{l}
      G^2_{\rm 2-loop} (\xi(T)), \quad T \ge T_c,
      \\[3mm]
      G^2_{\rm 2-loop}(\xi(T_c)) + b (1- T / T_c), \quad T < T_c ,
    \end{array}
  \right.
\end{equation}
where $G^2_{\rm 2-loop}$ is the relevant part of the 2-loop running coupling. Here, 
$\xi(T) = \lambda (T - T_s)/T_c$ contains a scale parameter $\lambda$ and an infrared regulator 
$T_s$. The effective coupling $G^2$ for arbitrary $T$ and $\mu_q$ can be found by solving a 
quasi-linear partial differential equation which follows from Maxwell's relation 
(cf.~\cite{Peshier,Bluhm04} for details of the model). In the next section, we want to test our phenomenological 
QPM by comparing with recent two-flavour ($N_f=2$) lattice QCD results \cite{Ejiri,Kar1,All05}. 
 
\section{Equation of State for $N_f=2$ \label{sec:EoS}}

For small $\mu_q$, the pressure can be expanded into a Taylor series in powers of $(\mu_q/T)$, 
\begin{equation}
  \label{e:presdecomp}
  p(T, \mu_q) = T^4 \sum_{n=0}^\infty c_n(T) \left( \frac{\mu_q}{T} \right)^n ,
\end{equation}
which has recently been studied in lattice QCD \cite{Kar1,All05}. The Taylor series was calculated 
up to including order $(\mu_q/T)^6$. $c_n(T)$, vanishing for odd $n$, follow straightforwardly from (\ref{e:pres0}) 
through differentiation, 
\begin{equation}
  \label{e:coeff1}
  c_n(T) = \left.\frac{1}{n!} \frac{\partial^n (p/T^4)}{\partial  (\mu_q/T)^n}\right|_{\mu_q = 0} \,.
\end{equation}

\begin{figure}[t]
\begin{minipage}{12.5cm}
  \centering{
  \includegraphics[scale=0.23,angle=-90.]{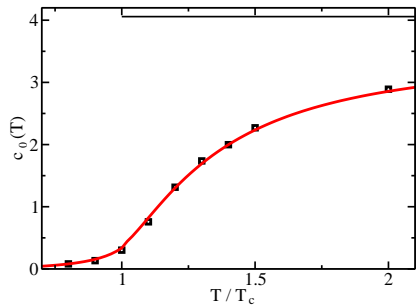}}
  \caption{Comparison of our QPM with lattice QCD results (symbols) for $c_0(T)$ 
    as function of $T/T_c$ for $N_f=2$. Raw lattice QCD data~\cite{Kar1} 
    are continuum extrapolated by an extrapolation factor $d=1.1$ for $T>T_c$ as advocated in 
    \cite{Kar1,Kar3} due to finite size and cut-off effects. 
    QPM parameters: $\lambda = 4.4$, $T_s=0.67T_c$, $b=344.4$, $B(T_c)=0.31 T_c^4$, setting 
    $T_c=175$ MeV as in \cite{Ejiri}. The horizontal line depicts the corresponding Stefan-Boltzmann value 
    highlighting the effects of strong interaction near $T_c$. \label{fig:2}}
\end{minipage}
\end{figure}

In Fig.~\ref{fig:2}, we compare QPM with lattice QCD results for $c_0(T)=p(T,\mu_q=0)/T^4$. 
In analogy to the lattice simulations, we set $m_{0;q}(T)=0.4T$. 
Furthermore, in \cite{Blu05} an impressively good agreement between the QPM results of $c_{2,4,6}(T)$ and the 
lattice QCD data was shown. 

From (\ref{e:presdecomp}), other thermodynamic quantities such as net baryon density $n_B=\partial p/\partial\mu_B$, 
entropy density $s$ and energy density $e$ follow as 
\begin{equation}
  \label{e:sn}
  s(T,\mu_B)  =  T^3\sum_{n=0}^\infty s_n(T)\left(\frac{\mu_B}{3T}\right)^n \,\,,\,\,\,\, 
  e(T,\mu_B)  =  T^4\sum_{n=0}^\infty e_n(T)\left(\frac{\mu_B}{3T}\right)^n , 
\end{equation}
where $e_n(T) = 3 c_n(T) + c_n'(T)$ and $s_n(T) = (4-n) c_n(T) + c_n'(T)$. Here, $\mu_B=3\mu_q$ denotes the 
baryo-chemical potential and $c_n'(T)=T dc_n(T)/dT$. 
Since $s_n(T)$ and $e_n(T)$ contain both, $c_n(T)$ and $c_n'(T)$, they serve for a more 
sensitive test of the model than considering $c_n(T)$ alone. 
\begin{figure}[t] 
\begin{minipage}{12.5cm}
  \includegraphics[scale=0.23,angle=-90.]{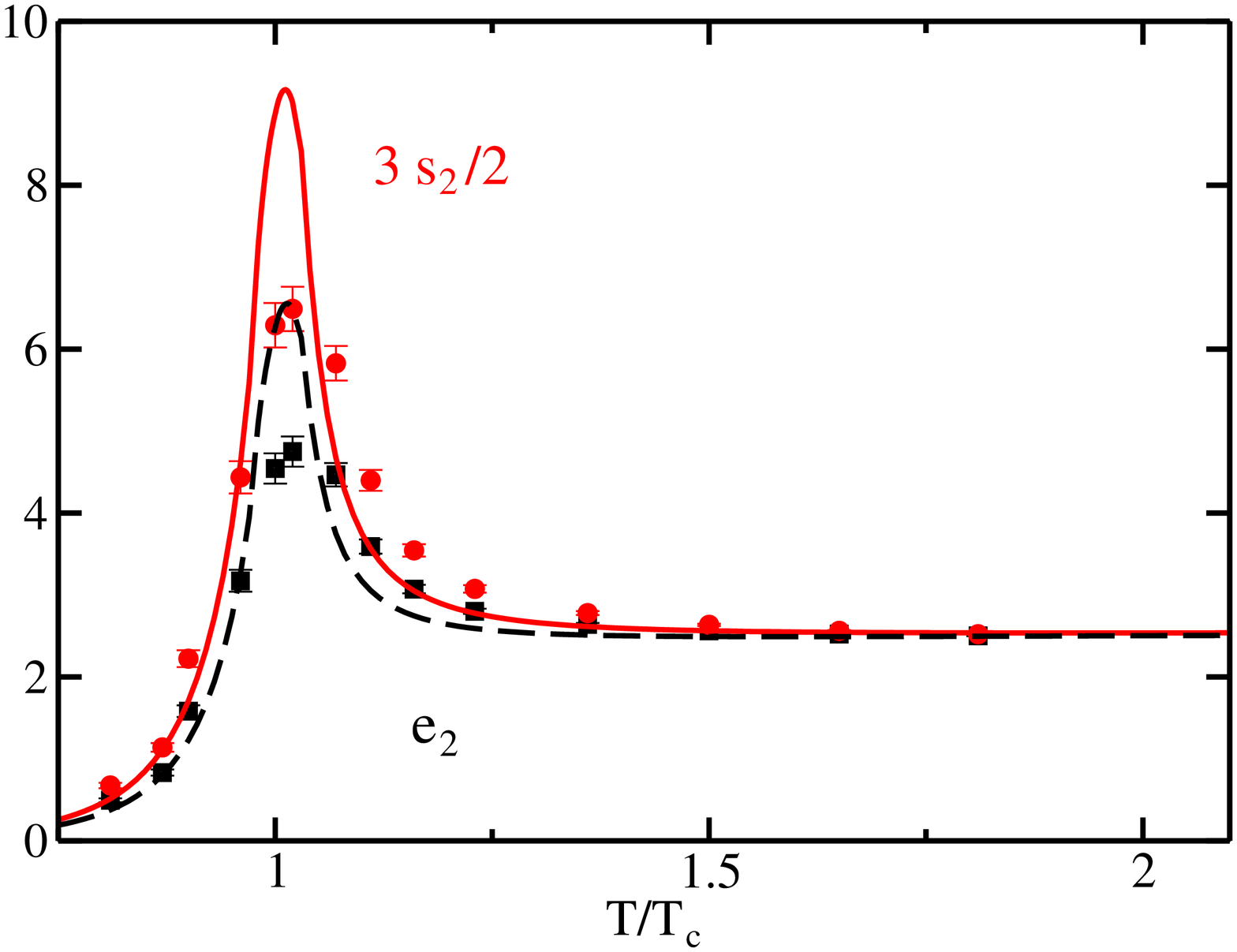}
  \includegraphics[scale=0.23,angle=-90.]{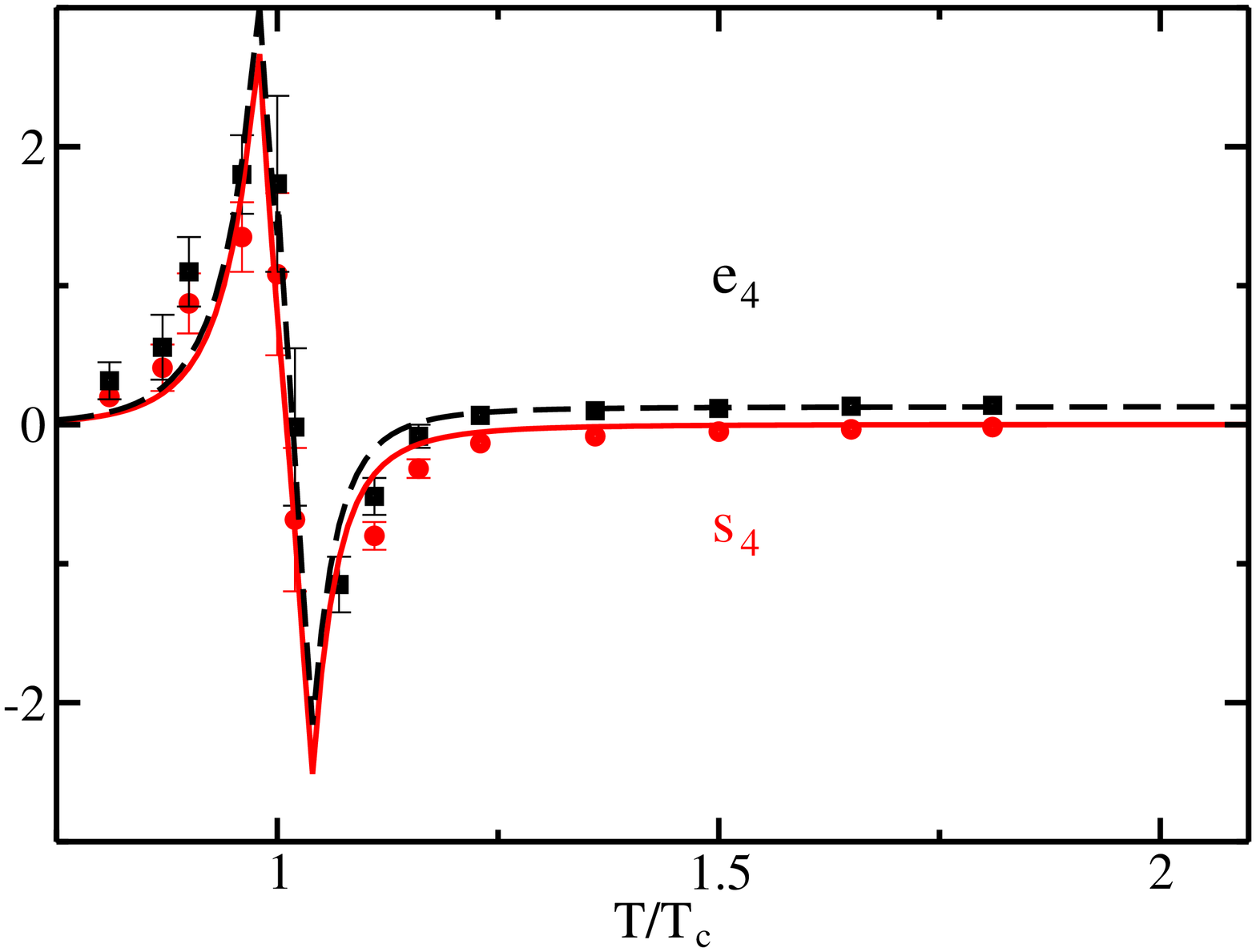}
  \caption{Comparison of our QPM with lattice QCD results \cite{Ejiri} for $e_n(T)$ (squares) and $s_n(T)$ (circles) 
    as function of $T/T_c$ for $N_f=2$; left (right) panel for $n=2\,(4)$. 
    QPM parameters: $\lambda = 12$, $T_s=0.87T_c$, $b=426.1$ with $T_c=175$ MeV 
    as adjusted to $c_2(T)$ from \cite{All05} (cf. \cite{Blu05}). We choose $\Delta T=5.25$ MeV for suitably estimating 
    $c_n'(T)$ by finite difference approximation of $c_n(T)$. \label{fig:coeffs2}}
\end{minipage}
\end{figure}
Estimating $c_n'(T)$ through a fine but finite difference approximation of $c_n(T)$, we compare QPM with lattice 
QCD results \cite{Ejiri} for $s_{2,4}$ 
and $e_{2,4}$ in Fig.~\ref{fig:coeffs2} and find a fairly good agreement. The pronounced structures in the vicinity 
of the transition temperature follow from the change in the curvature of $G^2(T,\mu_q=0)$ at $T=T_c$. 

\begin{figure}[b]
\begin{minipage}{12.5cm}
  \centering{
  \includegraphics[scale=0.23,angle=-90.]{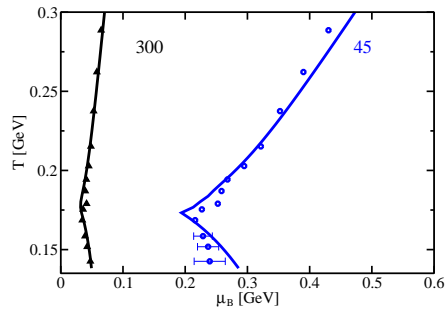}}
  \caption{Isentropic trajectories: Comparison of $N_f=2$ lattice QCD results \cite{Ejiri} for $s/n_B = $ 300 (triangles) 
    and 45 (circles) with the corresponding QPM results. The physical scale is $T_c=175$ MeV. \label{fig:isentrops}}
\end{minipage}
\end{figure}

Assuming local entropy and baryon number conservation during the hydrodynamical expansion of the fireball created 
in heavy-ion collisions, 
the strongly interacting system evolves isentropically. The evolutionary paths of individual 
fluid elements can be displayed by trajectories $s/n_B = const$ in the $T$ - $\mu_B$ plane. We calculate $n_B$ and 
$s$ from (\ref{e:sn}) up to $\mathcal{O}((\mu_B/T)^6)$ for the isentropic trajectories $s/n_B=$ 300, 45 
and compare with lattice QCD results \cite{Ejiri} in Fig.~\ref{fig:isentrops}. 
Finding a fairly good agreement, the pattern of the evolutionary paths is mainly influenced by $s_0(T)$. 
For instance, in the vicinity of $T_c$ a larger value of $s_0(T)$ of about 30\% translates into a 28\% increase 
in $\mu_B$ for the same trajectory. At large $T$, where $c_{0,2}(T)$ are 
essentially flat, the relation $\frac{\mu_B}{T} = 18 \frac{c_0}{c_2} (\frac{s}{n_B})^{-1}$ holds for small 
$\mu_B$, i.~e. lines of constant specific entropy are given by lines of constant $\mu_B/T$. 

\begin{figure}[b]
\begin{minipage}{12.5cm}
  \includegraphics[scale=0.23,angle=-90.]{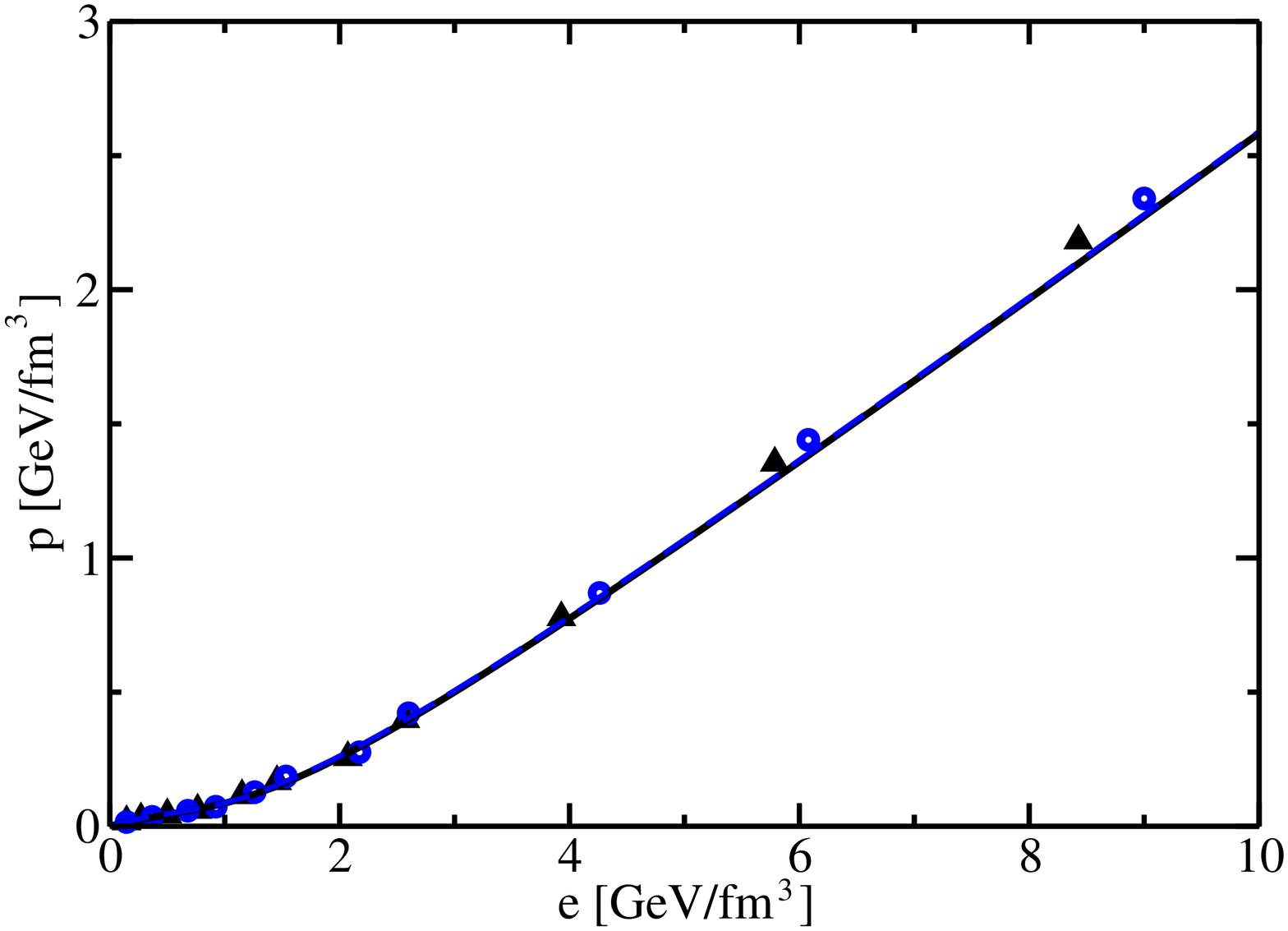}
  \includegraphics[scale=0.23,angle=-90.]{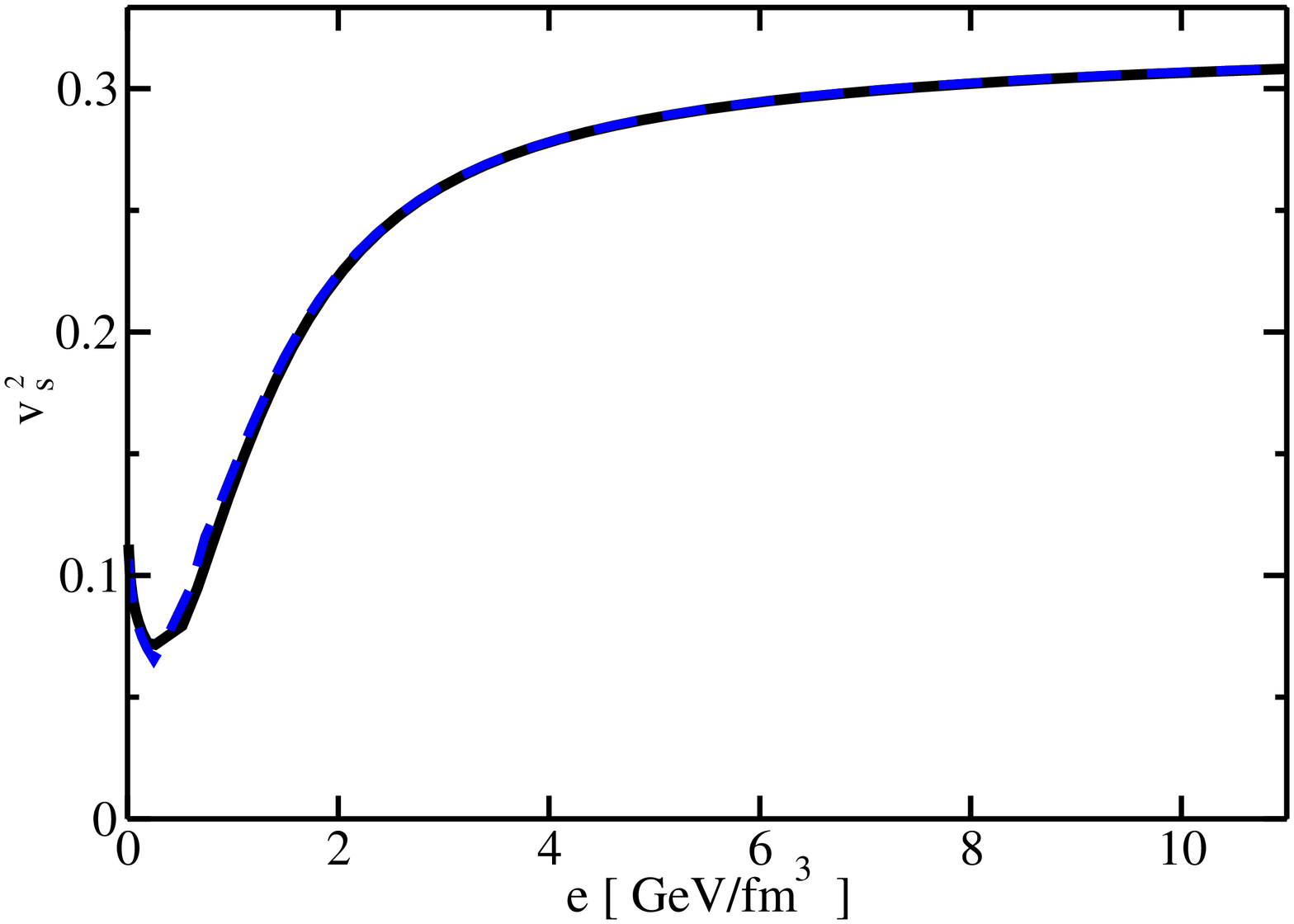}
  \caption{Left panel: comparison of lattice QCD results \cite{Ejiri} of $p$ as function of $e$ for $N_f=2$ along 
    $s/n_B=$ 300 (triangles) and 45 (circles) with corresponding QPM results depicted by solid and dashed 
    lines, respectively. Right panel: according speed of sound $v_s^2$. \label{fig:suppl1}}
\end{minipage}
\end{figure}

Along the isentropic trajectories, we evaluate the EoS $p(e)$. As depicted in Fig.~\ref{fig:suppl1} (left panel), 
the EoS, reproducing the lattice QCD results impressively well, is found to be almost independent 
of the considered specific entropy. 
Accordingly, the speed of sound $v_s^2=\partial p/\partial e$ as exhibited in the right panel of Fig.~\ref{fig:suppl1} is also 
rather independent of $s/n_B$. 
 
\section{Conclusions \label{sec:conclusions}} 

We presented a comparison of our QPM with recent two-flavour lattice QCD results of the isentropic equation of state at 
finite baryo-chemical potential. In particular, we focused on the Taylor expansion coefficients of energy density and 
entropy density, reproducing the pronounced structures in the vicinity of the (pseudo-) critical temperature fairly well. 
In addition, the isentropic trajectories in the $T$ - $\mu_B$ plane were compared. The EoS along those paths of constant 
entropy per baryon was found to be rather independent of the particular value of specific entropy. Having tested 
the successful applicability of our model in the finite $\mu_B$ - region for $N_f=2$, one can proceed to the physically 
interesting case of $N_f=2+1$ quark flavours and extend the EoS $p(T)$ towards finite baryon densities. Such an 
EoS can be applied to the hydrodynamical stage of heavy-ion collisions, e.~g. by studying transverse momentum spectra 
and differential elliptic flow of various hadron species. This will be reported elsewhere. 

\section*{Acknowledgment(s)}

This work is supported by BMBF 06 DR 121, GSI and Helmholtz association VI. We thank F. Karsch, 
E. Laermann, A. Peshier and K. Redlich for fruitful discussions. 
 

\vfill\eject
\end{document}